\documentclass[twocolumn]{aastex63}

\definecolor{PineGreen}{RGB}{1, 121, 111}

\newcommand\G{\textcolor{cyan}}

\let\oldFootnote\footnote
\newcommand\nextToken\relax

\renewcommand\footnote[1]{%
    \oldFootnote{#1}\futurelet\nextToken\isFootnote}

\newcommand\isFootnote{%
    \ifx\footnote\nextToken\textsuperscript{,}\fi}

\usepackage{CJK}
\usepackage{hyperref}
\usepackage{epsfig}
\usepackage{color}
\usepackage{ulem}

\shorttitle{LAMOST OBA STARS}
\shortauthors{Guo et al.}

\graphicspath{{./}{figures/}}

\begin{document}

\title{The Early-type Stars from LAMOST survey: Atmospheric parameters}

\correspondingauthor{Chao Liu; Xuefei Chen; Luqian Wang}
\email{liuchao@bao.ac.cn; cxf@ynao.ac.cn; wangluqian@ynao.ac.cn}

\author[0000-0001-9989-9834]{YanJun Guo}
\affiliation{Yunnan observatories, Chinese Academy of Sciences, P.O. Box 110, Kunming, 650011, China}
\affiliation{School of Astronomy and Space Science, University of Chinese Academy of Sciences, Beijing, 100049, People's Republic of China}
\affiliation{Key Laboratory for Structure and Evolution of Celestial Objects, Chinese Academy of Sciences, P.O. Box 110, Kunming 650216, People's Republic of China}

\author[0000-0002-6434-7201]{Bo Zhang}
\affiliation{Department of Astronomy, Beijing Normal University, Beijing 100875, People's Republic of China}

\author[0000-0002-1802-6917]{Chao Liu}
\affiliation{School of Astronomy and Space Science, University of Chinese Academy of Sciences, Beijing, 100049, People's Republic of China}
\affiliation{Key Laboratory of Space Astronomy and Technology, National Astronomical Observatories, Chinese Academy of Sciences, Beijing 100101, People's Republic of China}

\author[0000-0002-2577-1990]{Jiao Li}
\affiliation{Key Laboratory of Space Astronomy and Technology, National Astronomical Observatories, Chinese Academy of Sciences, Beijing 100101, People's Republic of China}

\author[0000-0003-3832-8864]{JiangDan Li}
\affiliation{Yunnan observatories, Chinese Academy of Sciences, P.O. Box 110, Kunming, 650011, China}
\affiliation{School of Astronomy and Space Science, University of Chinese Academy of Sciences, Beijing, 100049, People's Republic of China}

\author[0000-0003-4511-6800]{Luqian Wang}
\affiliation{Yunnan observatories, Chinese Academy of Sciences, P.O. Box 110, Kunming, 650011, China}

\author[0000-0001-5314-2924]{ZhiCun Liu}
\affiliation{University of Chinese Academy of Sciences Beijing 100049, People's Republic of China}

\author[0000-0002-3701-6626]{Yong-Hui Hou}
\affiliation{Nanjing Institute of Astronomical Optics \& Technology, National Astronomical Observatories, Chinese Academy of Sciences, Nanjing 210042, People’s Republic of China}
\affiliation{School of Astronomy and Space Science, University of Chinese Academy of Sciences, Beijing, 100049, People's Republic of China}

\author[0000-0001-9204-7778]{ZhanWen Han}
\affiliation{Yunnan observatories, Chinese Academy of Sciences, P.O. Box 110, Kunming, 650011, China}
\affiliation{School of Astronomy and Space Science, University of Chinese Academy of Sciences, Beijing, 100049, People's Republic of China}
\affiliation{Center for Astronomical Mega-Science, Chinese Academy of Sciences, Beijing 100012, China}

\author[0000-0001-5284-8001]{XueFei Chen}
\affiliation{Yunnan observatories, Chinese Academy of Sciences, P.O. Box 110, Kunming, 650011, China}
\affiliation{School of Astronomy and Space Science, University of Chinese Academy of Sciences, Beijing, 100049, People's Republic of China}
\affiliation{Center for Astronomical Mega-Science, Chinese Academy of Sciences, Beijing 100012, China}

\begin{abstract}
Massive stars play key roles in many astrophysical processes. Deriving atmospheric parameters of massive stars is important to understand their physical properties and thus are key inputs to trace their evolution. Here we report our work on adopting the data-driven technique Stellar LAbel Machine ({\tt SLAM}) with the non-LTE TLUSTY synthetic spectra as the training dataset to estimate the stellar parameters of LAMOST optical spectra for early-type stars. We apply two consistency tests to verify this machine learning method 
and compare stellar labels given by {\tt SLAM} with that in literature 
for several objects having high-resolution spectra. 
We provide the stellar labels of effective temperature ($T_\mathrm{eff}$), 
surface gravity ($\log{g}$), metallicity ([M/H]), and projected rotational velocity ($v\sin{i}$) 
for 3,931 and 578 early-type stars from LAMOST Low-Resolution Survey (LAMOST-LRS) and Medium-Resolution Survey (LAMOST-MRS), respectively. To estimate the average statistical uncertainties of our results, we calculated the standard deviation between the predicted stellar label and the pre-labeled published values from the high-resolution spectra. The uncertainties of the four parameters are $\sigma(T_\mathrm{eff}) = 2,185 $K, $\sigma(\log{g}) = 0.29$ dex, and $\sigma(v\sin{i}) = 11\, \rm km\,s^{-1}$ for MRS, and $\sigma(T_\mathrm{eff}) = 1,642 $K, $\sigma(\log{g}) = 0.25$ dex, and $\sigma(v\sin{i}) = 42\, \rm km\,s^{-1}$ for LRS spectra, respectively. We notice that parameters of $T_\mathrm{eff}$, $\log{g}$ and [M/H] can be better constrained 
using LRS spectra rather than using MRS spectra, most likely due to their broad wavelength coverage,
while $v\sin{i}$ is constrained better by MRS spectra than by LRS spectra, probably due to the relatively accurate line profiles of MRS spectra. 
\end{abstract}



\section{Introduction}\label{sec:intro}
Early-type stars are massive and luminous, and they are mainly comprised of O- or B-type stars \citep{1968Morton,1973Morgan,1973Panagia}. Massive early-type stars are possible progenitors of extremely compact stellar objects, such as 
neutron stars, black holes, high mass X-ray binaries, Type Ib/c supernovae, and are potential sources of gravitational wave events 
\citep{2008Sadowski,2010Yoon,2016GW1,2016GW2,2018XueFei,2020GWbinary,2020Hanzhanwen}. Stellar atmospheric parameters, such as effective temperature ($T_\mathrm{eff}$), surface gravity $\log{g}$, projected rotational velocity $v\sin{i}$, 
and metallicity [M/H]) are often referred to as the stellar labels. Deriving such labels for massive stars are important to reveal their physical properties 
and to constrain their location on the Hertzsprung-Russell (HR) diagram, and this information is an essential component to trace the evolutionary 
scenario of massive stars \citep{1990Fitzpatrick,1995Langer,1999McErlean}. Spectral chemical analysis of massive stars could be used
to determine the present-day abundance of local and external galaxies \citep{2001Daflon,urbaneja2005,2017Esteban}.
Early-type stars are also important for studies of cosmic ionization and acting as the major sources of energetic feedback for the interstellar medium (ISM) and intergalactic medium (IGM) \citep{hopkins2014}.

The classical way to derive stellar labels of massive stars is by comparing observations to theoretical stellar atmospheric models through $\chi^2$ 
minimization approaches. 
We summarize several works on determining the stellar labels of early-type stars using such techniques as follows. \citet{trundle2007} investigated 
stellar labels of 61 B-type stars in the Galaxy and Magellanic Clouds using the high-resolution spectra from the Very Large Telescope (VLT). 
\citet{hunter2009} obtained surface nitrogen abundance and projected rotational velocity for about 150 B-type stars located in the Galaxy and Small 
Magellanic Clouds using the spectroscopic observations from the VLT-FLAMES survey. A comprehensive study of the early B-type stars located 
within the solar neighborhood is carried out by \citet{nieva2012} using multiple sets of high-resolution spectra. \citet{mvevoy2017} measured 
atmospheric parameters and metallic abundance for a sample of runaway B-type stars in the Galaxy to trace their formation mechanism. 
These studies above all utilized high signal-to-noise ratio (SNR) spectra that cover the wavelength regions rich in atmospheric absorption lines to proceed
the line diagnostic analysis but the investigations were limited to a small sample of observations.

The alternative way to measure stellar labels for stellar systems is a data-driven approach. This technique learns the pre-labeled reference spectra and forms a model that maps the stellar labels to spectra. The data-driven 
technique is widely used and demonstrated reliable applicability in many astronomical studies. For example, \citet{2015Ness} constructed the ``The Cannon" model using a large set of observed spectra for reference stars with prior known stellar labels, and such model was applied to spectra 
observed from the APOGEE DR10 to estimate the stellar labels for about 55,000 stars. The predicted results are in good agreement with those 
obtained from the APOGEE pipeline. \cite{2020xiangmaoshengOBdistance} performed a neural network technique using the low-resolution spectra 
from Large Sky Area Multi-Object Fiber Spectroscopic Telescope (LAMOST) DR5 to estimate the absolute magnitude for a sample of 16,002 OB 
stars. \cite{2020zhangbolaspecslam} developed the data-driven learning module of Stellar LAbel Machine ({\it SLAM}) and applied it to spectra in the 
APOGEE DR15 spectral library to derive stellar labels for K-giant stars. More recently, \cite{2020Lijiadong} adopted the {\it SLAM} to 
derive stellar labels for M dwarfs using spectral observations from the LAMOST-MRS database.

There are only a few numbers of early-type star surveys available \citep{2010MaizApellaniz,2010Barba,2011SimonDiaz}. 
However, stellar labels derived using the spectra from these surveys vary from case to case, mainly because the spectral observations were taken on different instruments. Consequently, different pipelines and methodologies were applied to analyze the spectral properties of the data \citep{2017Sana}. Catalogs of stellar labels predicted from a large sample of homogeneous spectral observations using a consistent methodology are critically lacking. Motivated by the recent release of a large number of spectral observations from the LAMOST database, in this paper, we report our work of constructing a catalog consisting of homogenized stellar atmospheric parameters for early-type stars identified from the survey by applying the data-driven learning module {\tt SLAM}. Such catalog is the first one that consists of consistently derived stellar labels for such a large sample of early-type stars, and this information will be helpful to provide a reference for studying the physical and evolutionary properties of massive stars.

In the following of the text, we describe the sample selection of the LAMOST spectra in Section~\ref{sec:OB candidates}. 
We briefly describe the algorithm of the data-driven learning module {\tt SLAM}, the consistency check of applying the module to the observational data, and details of constructing the training dataset in Section~\ref{sec:slam}. We report our results of predicting stellar labels for the LAMOST spectra in Section~\ref{sec:results}. We discuss our results in Section~\ref{sec:Discussion} and summarize the conclusion in \ref{sec:Conclusion}.

\section{LAMOST DATA Selection} \label{sec:OB candidates}
LAMOST is a 4-meter quasi-meridian reflecting Schmidt telescope located at the Xinlong station of the National Astronomical Observatory. The telescope has a field of view of 20 square degrees and is implemented with 4000 fibers. Both low-resolution spectrograph (LRS) and medium-resolution spectrograph (MRS) were installed on the telescope. The LRS has a resolving power of $R\sim 1800$ with a wavelength coverage within a range of $3690\sim9100$~\AA, about $\sim$ 9 million\footnote{\url{http://dr5.lamost.org/}} of optical spectra are released at the time of writing \citep{2012CuiXiangQun,2012ZhaoGang,2012DengLiCai}. The MRS has $R\sim 7500$ and includes both a blue arm (B) and a red arm (R). Spectral observations made from the blue arm spanning wavelength in a range of $4950\sim5350$~\AA\, and the spectra observed using the red arm cover a wavelength range of $6300\sim6800$~\AA\ \citep{2020LiuChao}. LAMOST finished the first observing campaign from 2011 to 2018 (known as LAMOST I). The second active survey program (LAMOST II) began in 2017 September. A total number of about five million medium-resolution spectra observed with a single exposure for $\sim$800,000 stars were collected by 2019 June \citep{2020LiuChao}.

\begin{figure*}
	\centering
	\includegraphics[scale=0.7]{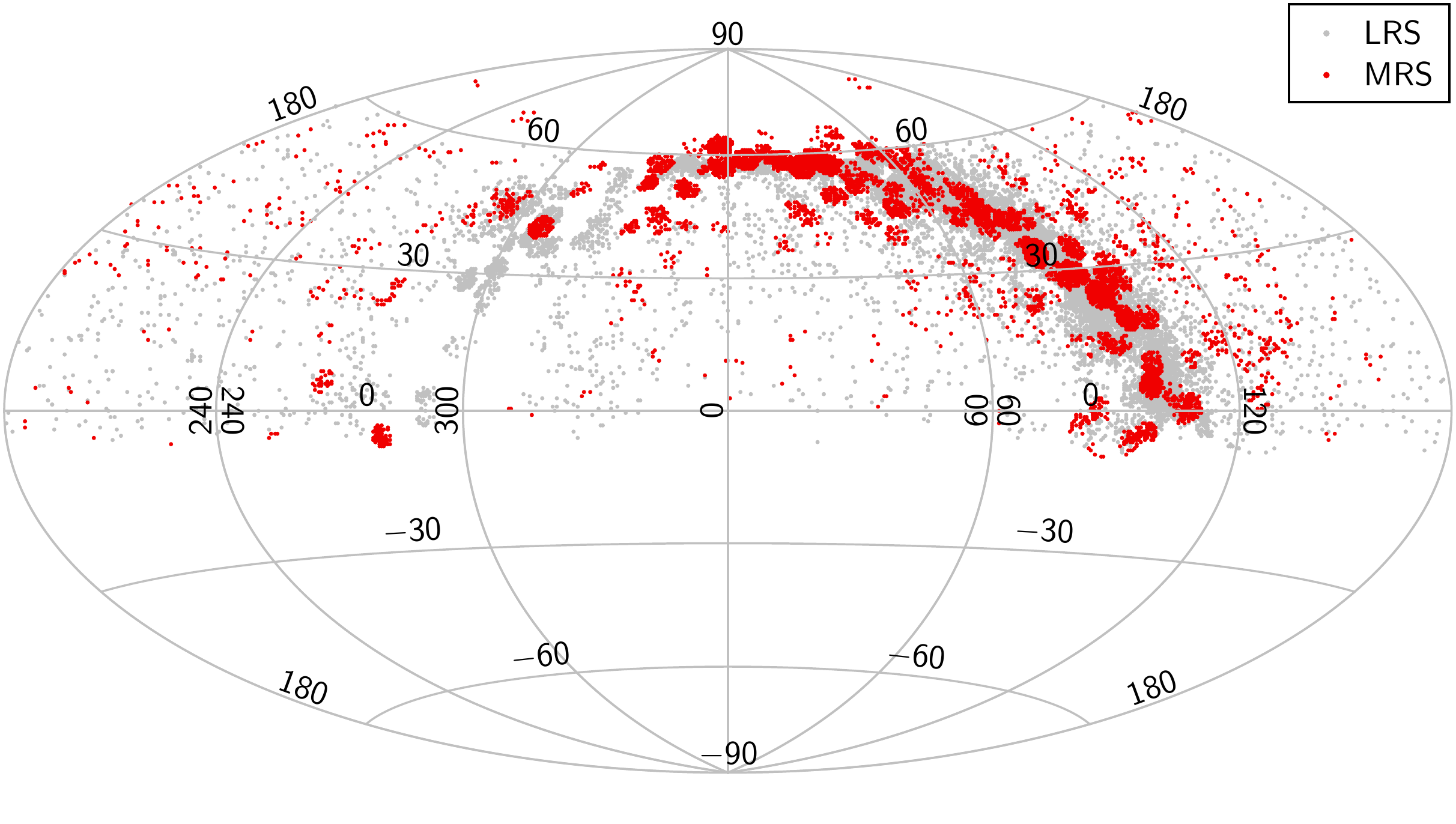}
    \caption{The spatial distribution of 16,032 early-type stars identified using observations from LAMOST-LRS (grey dots) by \cite{2019LiuZhicun} and 9,382 early-type stars identified using MRS observations (red dots) from LAMOST by \cite{2021guoguo}.}
    \label{fig: distribution}
\end{figure*}

\cite{2019LiuZhicun} identified 16,032 early-type stars through measuring the equivalent widths of several absorption line profiles using low-resolution  spectra from the LAMOST LRS database. \cite{2021guoguo} utilized the spectra from LAMOST MRS database to investigate the multiplicity properties of early-type stars (hereinafter Paper I). In that work, a total number of 9,382 early-type stars were identified by measuring the equivalent widths of H$\alpha$ $\lambda$6564 \AA, He I $\lambda$6678 \AA, and Mg I profiles in the range of $5167\sim5184$~\AA. We display the distribution of the two samples in Figure~\ref{fig: distribution} in which the grey dots represent early-type stars from LAMOST-LRS, and red dots form LAMOST-MRS.
In this study, we adopt all of the identified O- and B-type stars from both \cite{2019LiuZhicun} and Paper~I to derive the stellar labels of their spectra by applying the {\tt SLAM}. 

\section{Method} \label{sec:slam}
\subsection{SLAM}\label{sec:training process}
The Stellar LAbel Machine ({\tt SLAM}) \citep{2020zhangbolaspecslam, 2020zhangboMRS} is a forward stellar spectral model based on the Support Vector Regression (SVR) algorithm \citep{cortes1995support}, which is widely applied in spectral analysis. A comprehensive discussion on the algorithm and architecture of the {\tt SLAM} are described in \citep{2020zhangbolaspecslam}. The {\tt SLAM} code is written in {\tt python}, and is available for public downloading from the GitHub website\footnote{\url{https://github.com/hypergravity/astroslam}}.
In order to apply the {\tt SLAM} to observed spectral data, we adopted several default hyperparameters as suggested by \citet{2020zhangbolaspecslam} to describe the model complexities in the SVR algorithm. These include the penalty level ($C$) within a range of [0.1,1,10], the width of the radial basis function ($\gamma$) under the range of [0.1,1], and tube radius ($\epsilon$) with a value of 0.05.

In principle, three steps are included for applying the {\tt SLAM}  to observational spectra data, and we briefly summarize these as follows: \newline
Step 1: Pre-processing. We first need to standardize the spectra in the training set such that their stellar labels and spectral fluxes have a mean value of 0 and variance with a value of 1. This step aims to map the stellar labels of spectra in the training set and normalize the spectra into standardized space. This procedure is automatically achieved by {\tt SLAM} \citep{2020zhangbolaspecslam} through an internal computational algorithm. \newline
Step 2: Training. We use the spectra in the training set to train the SVR model such that it learns the knowledge of the pre-labeled spectra from input spectral data themselves, and then it maps the corresponding atmospheric parameters and spectra to a model. \newline
Step 3: Prediction. Based upon the training procedure as described above, we predict the stellar labels for observed spectral data by using the SVR model.
\subsection{The training set}\label{sec:train set}
\begin{figure}
	\centering
	\includegraphics[scale=0.7]{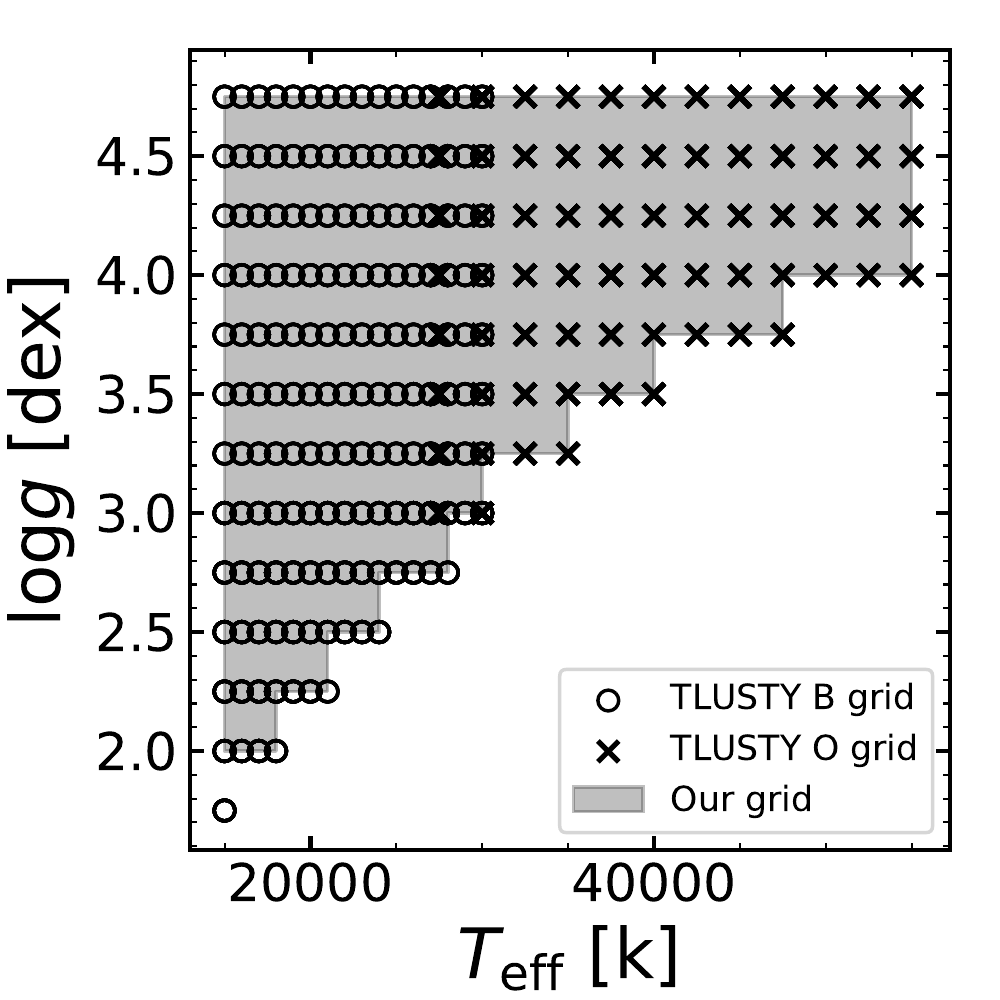}
    \caption{The TLUSTY atmospheric model grids for O-type stars (crosses) and B-type stars (circles) shown in the ($T_\mathrm{eff}$, $\log{g}$) plane. The grey shaded area denotes the linearly interpolated model grids used for the training set.}\label{fig:grid}
\end{figure} 

In order to apply the {\tt SLAM}  to spectra of early-type stars identified from the LAMOST database, we first need to construct a training set. 
We search in literature to collect OB stars with derived stellar labels, but the samples are inadequate to provide complete coverage of parameter space needed for stellar labels.
Therefore, we adopt the TLUSTY, a synthetic spectral library based on the non-Local Thermal Equilibrium (NLTE) model as our training set \citep{2003LanztlustyO,2007LanztlustyB}. The NLTE model includes model grids for both O- and B-type stars, including more than 40 elements in the calculations, such as the major atomic species of H, He, C, N, O, Ne, Si, P, S, Fe, and Ni \citep{2003LanztlustyO,2007LanztlustyB}. The solar abundances from \cite{1998Grevesse} are adopted in the model. The O-type model spectra assume a micro-turbulent velocity ($V_t$) of 10 $\rm km\,s^{-1}$, the spectra were sampled irregularly with a sample size of 180,000 $\sim$ 200,000 pixels, and the wavelength spans from 3,000 to 7,500 \AA. For B-type model spectra, about 175,000 pixel points were used to sample the model grids in irregular format with $V_t = 10$ $\rm km\,s^{-1}$. In addition, a $V_t = 2$ $\rm km\,s^{-1}$ was introduced to the spectra, and the model grids were sampled with 380,000 points. The B-type model grids cover a wavelength range of 3,200$\sim$10,000 \AA.

We adopted all the available O- and B-type model spectra from the grids. The O-type model grids include the $ T_\mathrm{eff}$ spanning a range of 27,500$\sim$55,000 K with a step size of 2500 K, the $\log{g}$ within a range of 3.00$\sim$4.75 dex with a step size of 0.25 dex. We chose the $T_\mathrm{eff}$ within a range of 15,000$\sim$30,000 K with a step size of 1000 K, the $\log{g}$ under a range of 1.75$\sim$4.75 dex with a step size of 0.25 dex for B-type stars. The metallicity ([M/H]) with values between $-1.0$ to 0.3 dex were chosen for both O-type and B-type model spectra. Projected rotational velocity and macro-turbulent velocities were not included in all of these spectra.
In Figure \ref{fig:grid}, we show the TLUSTY model grids for both O-type stars (crosses) and B-type stars (circles) in the ($T_\mathrm{eff}$, $\log{g}$) plane.

We downloaded the synthetic model O-type spectra\footnote{\url{http://tlusty.oca.eu/Tlusty2002/tlusty-frames-OS02.html}} and the B-type spectra\footnote{\url{http://tlusty.oca.eu/Tlusty2002/tlusty-frames-BS06.html}} from the TLUSTY websites. In order to increase the sample size, we randomly generated the model spectra using stellar parameters within the available range given by the TLUSTY grids through a linear interpolation approach, and these expanded grids are shown as the gray shaded area on Figure~\ref{fig:grid}. We then randomly convolved these model spectra with projected rotational velocity $v\sin{i}$ with values within the range of 0 to 300\,$\rm km\,s^{-1}$. In order to bring the generated model spectra into an agreement with those of observations obtained from the LAMOST, we then degraded the resolutions of TLUSTY model spectra down to $R\sim1,800$ (comparable to LAMOST-LRS, defined as TLUSTY-LRS) and $R\sim7,500$ (comparable to LAMOST-MRS, defined as TLUSTY-MRS) through a Gaussian smoothing using the python package {\tt{laspec.qconv}}\footnote{\url{https://github.com/hypergravity/laspec/}}. We then re-sampled the wavelength grids of both TLUSTY-LRS and TLUSTY-MRS spectra such that they have comparable coverage as of observations, the wavelength spans a range of 3,900$\sim$7,000 \AA\ with a step size of 1 \AA\ for TLUSTY-LRS spectra, 4,950$\sim$5,350 \AA\ and 6,300$\sim$6,800 \AA\ for the red and blue arms of the TLUSTY-MRS spectra with a step size of 0.2 \AA, respectively. We excluded bad pixels from the spectra, and the training set consists of 1,000 TLUSTY-LRS spectra and 5,000 TLUSTY-MRS spectra (See Sec.~\ref{sec:err SNR}).

\subsection{Self-consistency check}\label{sec:consistency check}
Before we apply the {\tt SLAM} to the identified early-type stars from the LAMOST database to predict their stellar labels, we first need to verify the robustness of applying the module to the training set. The verification is achieved through the usage of a consistency check, Cross-Validation (CV). CV, is also called $k$-fold cross-validation, is a popular methodology used to score the performance of machine learning models in a less biased or less optimistic approach. 
This approach divides the input spectra into $k$ groups. Among them, the $k-1$ groups are randomly selected as the training set, and the spectra that remained in the $k$ set are chosen to be the validation set, such that stellar labels of spectra in the validation set will be predicted by using the {\tt SLAM}.

We adopted a 5-fold Cross-Validation technique from \cite{2010Ojala}. We use CV-Scatter (standard deviation) and CV-bias (mean deviation) to describe the performance of CV, and the equations are given as follows:

\begin{equation}\label{CVscatter}
    {\rm CV-scatter} = \frac{1}{n} \sqrt{\sum_{i=1}^{n}(\theta_{i,SLAM} - \theta_{i})^2},
\end{equation}
and
\begin{equation}\label{CVbias}
    {\rm CV-bias} = \frac{1}{n} \sum_{i=1}^{n}(\theta_{i,SLAM} - \theta_{i}),
\end{equation}

where $\theta_{i}$ is the true stellar label of the star with an index of i. $\theta_{i,SLAM}$ denotes the stellar label predicted from the {\tt SLAM} of the same star.

The values of CV-Scatter and CV-bias are dependent on the SNR of the input spectra. We thus investigate the distribution of the CV values as a function of the SNR added to the spectra. We added a set of SNR values to each pixel of the TLUSTY-LRS/MRS spectra using a Gaussian function to evaluate the performances of {\tt SLAM} in different SNRs. The SNR values span a range of $20\sim100$ with an increment of 20.

\begin{figure*}
	\centering
	\includegraphics[scale=0.5]{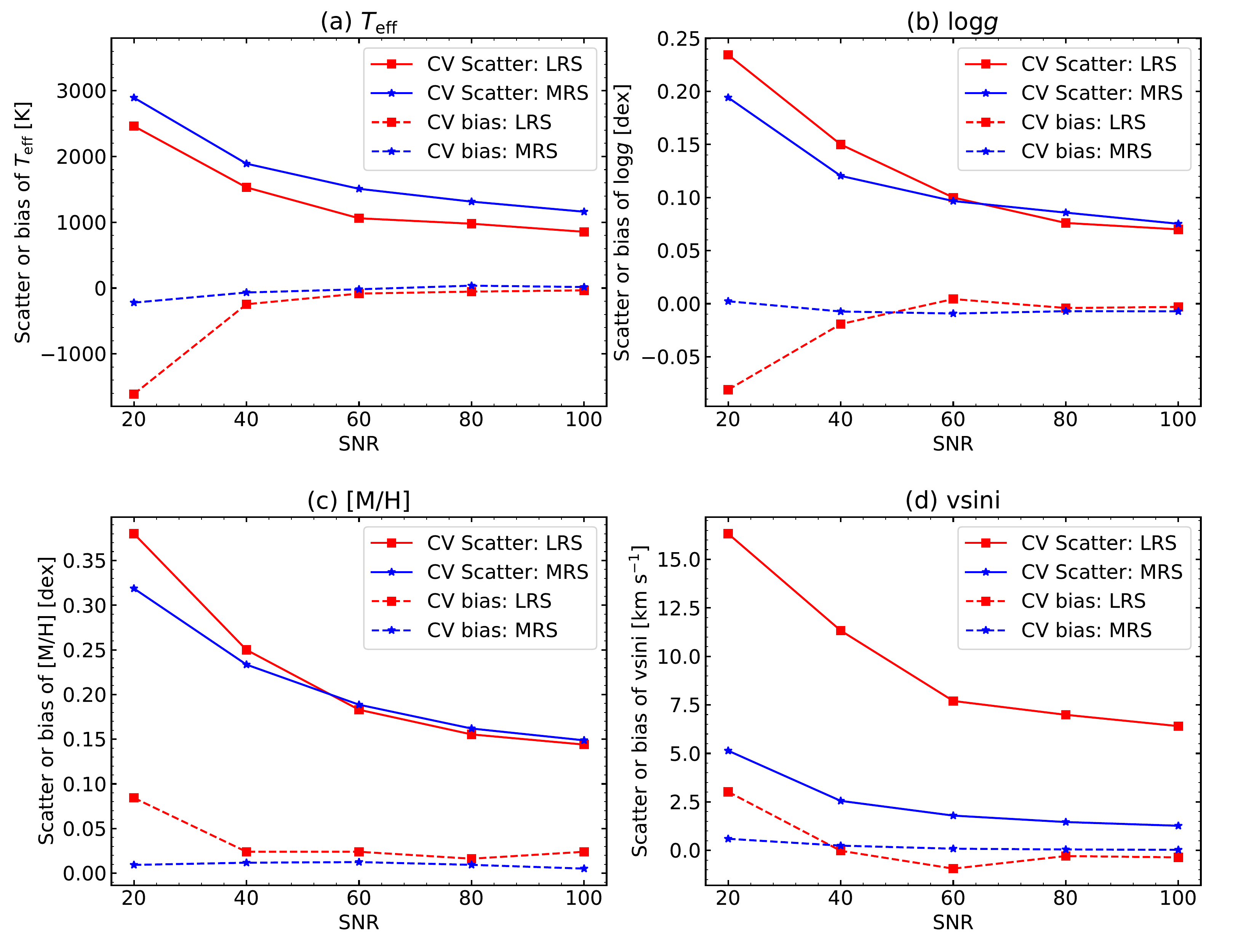}
    \caption{The distributions of the CV-scatter (solid lines) and the CV-bias (dash lines) values for predicted stellar labels of the $T_{\mathrm eff}$ (panel a), the $\log{g}$ (panel b), the $[M/H]$ (panel c), and $v\sin{i}$ (panel d) as a function of SNR values. Results obtained from the TLUSTY-LRS spectra and the TLUSTY-MRS spectra in the training set were shown in square and star symbols, respectively.}\label{fig:CV scatter bias}
\end{figure*}

In principle, a decreasing trend is shown from the CV bias, and CV scatter tests indicating that the data-driven method is suitable. In Fig.~\ref{fig:CV scatter bias}, we show the distributions for both the CV-Scatter and the CV-bias values for each predicted stellar label as a function of SNR. In panels (a) to (d), the CV scatter values calculated for $T_\mathrm{eff}$, $\log{g}$, $[M/H]$ and $v\sin{i}$ parameters using the TLUSTY-LRS (red lines with square symbols) and TLUSTY-MRS (blue lines with star symbols) spectra all display a decreasing trend towards a higher SNR. The calculated CV-bias values in panels (a) to (d) all approach to a zero towards higher SNR values, indicating that the stellar labels predicted by the {\tt SLAM} for spectra in the validation set are consistently matched with the pre-labeled values as given from the TLUSTY model grids.

\subsection{Determining the sample sizes and SLAM errors}\label{sec:err SNR}
\begin{figure*}
	\centering
	\includegraphics[scale=0.6]{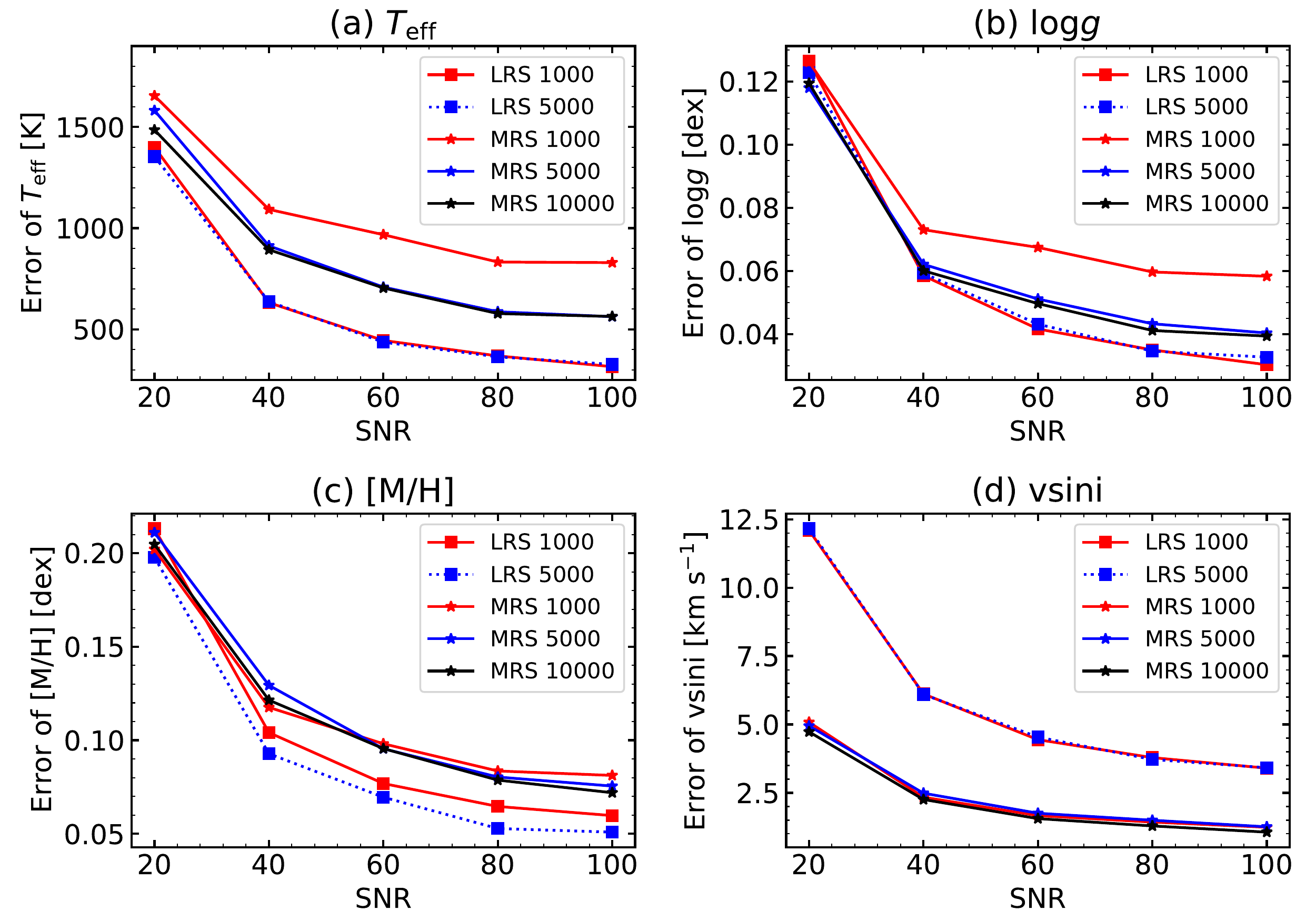}
    \caption{The distribution of SLAM errors of the predicted $T_\mathrm{eff}$ (panel a), $\log{g}$ (panel b), $[M/H]$ (panel c), and $v\sin{i}$ (panel d) plotted as a function of SNR values. The errors of stellar labels for TLUSTY-LRS spectra with a training sample size of 1,000 spectra (solid red lines with square symbols) and a set with 5,000 model spectra (dashed blue lines with square symbols) are shown in all panels. The errors of stellar labels for TLUSTY-MRS spectra with sample sizes of 1,000 (solid red lines with star symbols), 5,000 (solid blue lines with star symbols), and 10,000 spectra (solid black lines with star symbols) are shown in four panels.}
\label{fig:err SNR}
\end{figure*}

In principle, enlarging the size of the training sample by including more spectra results in smaller uncertainties of stellar labels predicted from the {\tt SLAM}, but the computation is costed by a significant amount of time. We thus need to determine the size of the training sample such that the precision of the prediction task can be compromised with a reasonable computing time budget. In addition to the original training set consisting of 1,000 TLUSTY-LRS model spectra, we constructed an extra
training set including 5,000 model spectra. Besides the exiting training set of 5,000 TLUSTY-MRS model spectra, we generated two additional sets consisting of 1,000 and 10,000 model spectra. By following the procedures as described in Sec.~\ref{sec:consistency check}, we added the set of SNR values to each pixel of the model spectra in individual training set for both TLUSTY-LRS and TLUSTY-MRS and then predicted their associated stellar labels. The {\tt SLAM} errors for each of the predicted stellar labels are obtained by computing the standard deviation between the predicted values from the {\tt SLAM} and the true labels as given from the TLUSTY model grids.

In Fig.~\ref{fig:err SNR}, we show the {\tt SLAM} errors of the predicted stellar labels as a function of SNR for both TLUSTY-LRS (lines with square symbols) and TLUSTY-MRS (lines with star symbols) spectra with different choices of sample sizes. For the two training sets of TLUSTY-LRS model spectra, including a collection of 1,000 (solid red lines with square symbols) and 5,000 (dashed blue lines with square symbols) spectra, the {\tt SLAM} errors of the predicted stellar labels all display a decreasing trend towards higher SNR values (panels a to d). The {\tt SLAM} errors of predicted stellar labels for these two sets become indistinguishable, and this suggests that the training set with a sample size of 1,000 is sufficient to maintain the precision of the prediction task. The {\tt SLAM} errors of the predicted stellar labels for the TLUSTY-MRS spectra display a similar decreasing trend as the TLUSTY-LRS. In panels (a) and (b) of Fig.~\ref{fig:err SNR}, significant deviations in the {\tt SLAM} errors of the predicted $T_\mathrm{eff}$ and $\log{g}$ for TLUSTY-MRS model spectra in training set with a sample size of 1,000 spectra (solid red lines with star symbols) are shown from the errors of the sets with sizes of 5,000 (solid blue lines with star symbols) and 10,000 spectra (solid black lines with star symbols). Such deviation of {\tt SLAM} errors indicates that the training set with a sample size of 1,000 spectra is inadequate to maintain a precise estimation for these stellar parameters. The {\tt SLAM} errors of the predicted $[M/H]$ and $v\sin{i}$ parameters for TLUSTY-MRS spectra in training sets with the three chosen sample sizes shown in panels (c) and (d) all converge to a similar small value towards higher SNR, and no significant variations are shown from the different sample sizes. We thus chose a sample size of 5,000 spectra for the TLUSTY-MRS training set.

\section{Results and Validation}\label{sec:results}
\begin{figure*}
	\centering
	\includegraphics[scale=0.7]{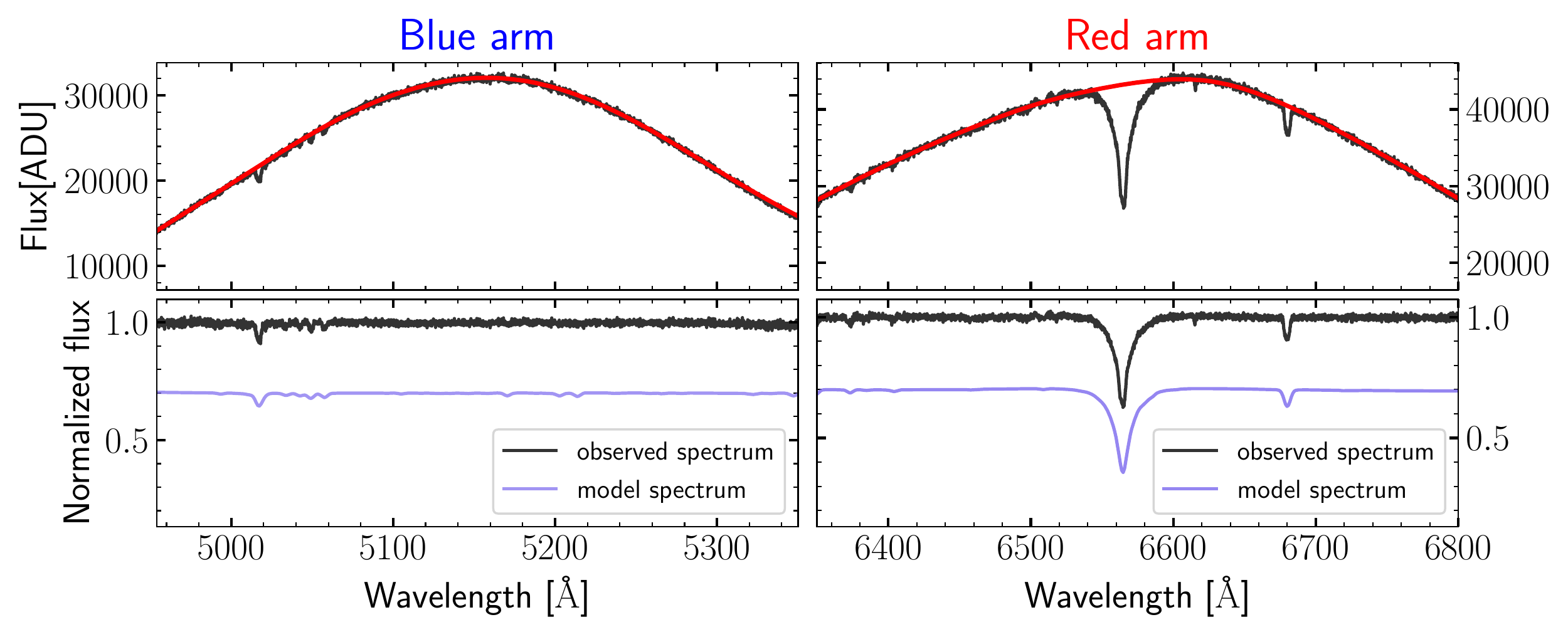}
	\caption{The upper panels display the LAMOST-MRS spectrum of OBSID 703901108 observed on night of MJD 58470.9540 (black lines) from the blue arm (left panel) and the red arm (right panel). A spline fit (red lines) was applied to the spectrum to perform the spectral normalization. The bottom panels show the normalized and RV corrected spectra (black lines) for both the blue arm (left panel) and the red arm (right panel). A model spectrum (purple lines) is over plotted with predicted stellar labels of $T_\mathrm{eff}=16841$K, $\log{g}= 4.3$ dex, [M/H]= -0.18 dex and $v\sin{i}= 111$ $\rm km\,s^{-1}$ obtained from the {\tt SLAM}.}
	\label{fig:MRS norm}
\end{figure*}

\begin{figure*}
	\centering
	\includegraphics[scale=0.7]{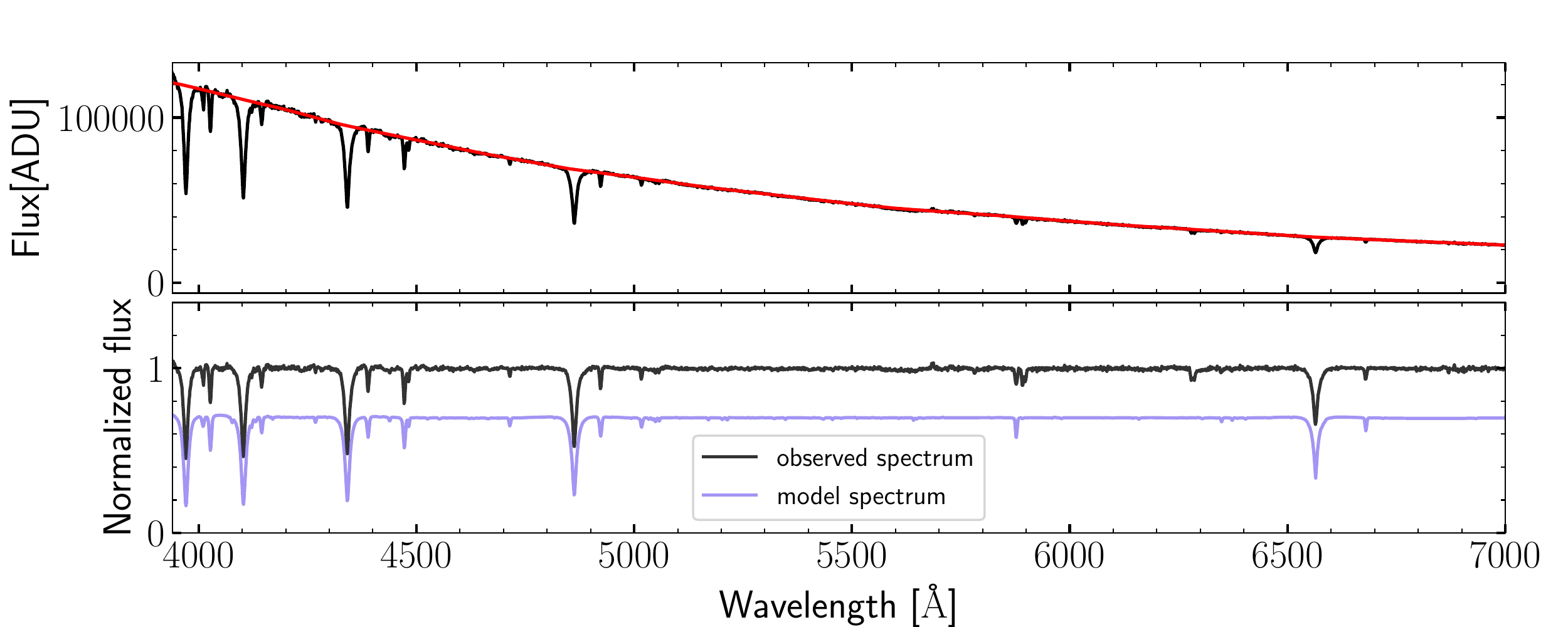}
	\caption{The upper panels display the LAMOST-LRS spectrum of the same star as shown in Fig.~\ref{fig:MRS norm} (black lines). A spline fit (red lines) was applied to the spectrum to perform the spectral normalization. The bottom panels show the normalized and RV corrected spectra (black lines). A model spectrum (purple lines) is over plotted with predicted stellar labels of $T_\mathrm{eff} = 16,494$K, $\log{g} = 4.2$ dex, $[M/H ]= -0.30$ dex, and $v\sin{i}= 4$ $\rm km\ s^{-1}$ obtained from the {\tt SLAM}.}
	\label{fig:LRS norm}
\end{figure*}

After learning the spectral features using the model spectra of TLUSTY-LRS and TLUSTY-MRS, now we are ready to apply the {\tt SLAM} to observational spectra for early-type stars from the LAMOST database to predict their stellar labels. We downloaded a collection of 16,032 LAMOST-LRS spectra and 9,382 LAMOST-MRS early-type star spectra from the LAMOST archive sites\footnote{\url{http://dr5.lamost.org/}}\footnote{\url{http://dr7.lamost.org/}}. The archived spectra were reduced, and wavelength calibration was applied following the pipeline as described in \citet{luo2015}. We then used the python package {\tt{laspec}}\footnote{\url{https://github.com/hypergravity/laspec}} to normalize the spectra using a spline fit\footnote{We use a smoothing spline to fit the observed spectrum and clip the pixels 3$\sigma$ above or 1$\sigma$ below the median of residual and iterate this process three times, to avoid strong lines, which is equivalent to a low pass filtering. We obtain the pseudo-continuum from the remaining smoothed pixels, and then the obtained normalized spectrum by dividing the observed spectrum by the pseudo-continuum.}. An example of LAMOST-MRS spectrum (black lines) with OBSID of 703901108 and the associated spline fit (red lines) are shown in upper panels of Fig.~\ref{fig:MRS norm}. The normalization procedure for the same star but with LAMOST-LRS observation is shown in the upper panel of Fig.~\ref{fig:LRS norm}. Before applying the {\tt SLAM} to the observational data, we first need to bring the LAMOST spectra into the rest frame such that they are in agreement with those of the TLUSTY model spectra. We achieved this goal by collecting the Radial Velocity (RV) measurements of our sample stars appearing in the work of \cite{2021zhangboRV}, who performed the RV measurements for all stars in the LAMOST DR7 through a cross-correlation algorithm. We then corrected the RV variations in our observed spectra. In order to avoid confusion in analyzing the spectra line profiles using the {\tt SLAM}, we omitted broad diffuse interstellar bands of 4430~\AA, 5780~\AA, 6196~\AA, 6283~\AA, and 6614~\AA\ \citep{1995HerbigIDBs}. We also excluded a gap appearing in the LAMOST-LRS spectra from 5800~\AA\ to 6000~\AA\ between the blue and red arms. In the bottom panels of Fig.~\ref{fig:MRS norm}, we show the normalized and RV corrected spectra (black lines). Based upon the stellar labels predicted from the {\tt SLAM}, this LAMOST-MRS spectrum has estimated $T_\mathrm{eff} = 16,841$ K, $\log{g} = 4.3$ dex, $[M/H] = -0.18$ dex, and $v\sin{i}= 111 \rm km\ s^{-1}$. We overplotted the model spectrum with the predicted stellar labels in purple lines in Fig.~\ref{fig:MRS norm}. The LAMOST-LRS spectrum of the same star has predicted stellar labels of $T_\mathrm{eff} = 16,494$ K, $\log{g} = 4.2$ dex, $[M/H ]= -0.30$ dex, and $
v\sin{i}= 4 \rm km\ s^{-1}$ as given by {\tt SLAM}, and the model spectrum with the estimated stellar labels are shown in Fig.~\ref{fig:LRS norm}.

\subsection{The stellar parameters of OB stars}
By applying the {\tt SLAM} to the sample of stars selected from the LAMOST database, we predict their stellar labels. In Table~\ref{Tab:1}, we report the spectral observational ID, the equatorial coordinates, SNR values, the predicted values of $T_\mathrm{eff}$, the $\log{g}$, the $[M/H]$, and $v\sin{i}$ of 16,032 early-type stars from LAMOST-LRS. Among the sample, the stellar labels of 3,931 stars are within the parameter ranges given by the training set. We assigned a flag of `I' to those stars, and they are listed in column 9 of Table~\ref{Tab:1}. For the remaining stars outside the parameter range, we obtained their stellar labels by extrapolating values from the {\tt SLAM}. We retained the predicted stellar labels of these stars in Table~\ref{Tab:1} as preliminary estimates for the stellar parameters. In Table~\ref{Tab:2}, we list the observational ID, the equatorial coordinates, SNR values, the predicted stellar labels of $T_\mathrm{eff}$, the $\log{g}$, the $[M/H]$, and $v\sin{i}$, the observational date (MJD), SNR\_R (red arm), SNR\_B (blue arm), and index of flags for 9,238 early-type stars from the LAMOST-MRS database. Among the sample, 578 stars show predicted stellar labels within the parameter ranges of the associated training set.

\begin{deluxetable*}{lcccccccccl}
\tablenum{1}
\tablecaption{The predicted stellar labels of 16,032 LAMOST-LRS early-type stars. \label{Tab:1}}
\tablewidth{0pt}
\tablehead{
\colhead{Star} & \colhead{RA} & \colhead{DEC} & \colhead{SNR} & \colhead{$T_\mathrm{eff}$} & \colhead{$\log{g}$}  & \colhead{$[M/H]$}  &  \colhead{$v\sin{i}$} & \colhead{Star}  \\
\colhead{OBSID} & \colhead{(deg)} & \colhead{(deg)} & \colhead{} & \colhead{(K)} & \colhead{(dex)}  & \colhead{(dex)}  &  \colhead{(km s$^{-1}$)} & \colhead{Flag}
}
\startdata
51401075  & 237.0048   		& \phn$-$2.5811 & 41  & 39714 & 4.4 & $-$0.74 & \phn6  	& I \\
203808205 & \phn93.7223   	& \phs12.3566 	& 91  & 31784 & 4.2 & $-$0.60 & \phn8  	& I \\
253216223 & 322.9148  		& \phs50.8778 	& 21  & 37341 & 4.0 & $-$0.56 & \phn1  	& I \\
255008029 & \phn81.8724   	& \phs36.4042 	& 45  & 18063 & 4.1 & $-$0.48 & \phn3  	& I \\
360106063 & 297.1958  		& \phs44.3561 	& 31  & 24675 & 4.3 & $-$0.09 & 13 		& I \\
536607134 & \phn95.1156  	& \phs21.8162 	& 51  & 15550 & 4.3 & $-$0.13 & \phn5  	& I \\
\enddata
\tablecomments{This table is available in its entirety in machine-readable form. The first six entires are shown here for guidance regarding
its format and content.}
\end{deluxetable*}

\begin{deluxetable*}{lccccccccclccr}
\tablenum{2}
\tablecaption{The predicted stellar parameters of 9,238  LAMOST-MRS early-type stars. \label{Tab:2}}
\tablewidth{0pt}
\tablehead{
\colhead{Star} & \colhead{RA} & \colhead{DEC} & \colhead{Date}  & \colhead{SNR$_{R}$} & \colhead{SNR$_{B}$}  & \colhead{$T_\mathrm{eff}$} & \colhead{$\log{g}$}  & \colhead{$[M/H]$}  &  \colhead{$v\sin{i}$} & \colhead{Star}  \\
\colhead{OBSID} & \colhead{(deg)} & \colhead{(deg)} & \colhead{(MJD)} & \colhead{} & \colhead{} & \colhead{(K)} & \colhead{(dex)}  & \colhead{(dex)}  &  \colhead{(km s$^{-1}$)} & \colhead{Flag}
}
\startdata
684408119& \phn35.4066  & 57.2101 & 58421.0472 & \phn97  	& \phn80  	& 25644 & 4.6 & $-$0.25 &112 	& I\\
624715119& \phn36.1931  & 57.4642 & 58119.8188 & 272 		& 190 		& 22287 & 2.8 & $-$0.75 &131 	& I\\
692513076& \phn48.8320  & 65.9180 & 58444.9562 & 101 		& \phn69  	& 17908 & 4.4 & $-$0.25 &\phn75 & I\\
729206077& \phn84.0657  & 34.2383 & 58539.8562 & \phn97  	& \phn56  	& 26271 & 4.4 & $-$0.09 &\phn36 & I\\
609116071& \phn18.4606 	& 59.7916 & 58088.8562 & \phn95   & 75        & 22445 & 3.8 & $-$0.10 &\phn5  & I\\
591515050& \phn19.3419 	& 58.8882 & 58030.9861  & 159  	& 131       & 24700 & 3.7 & $-$0.10 &\phn44 & I\\
\enddata
\tablecomments{This table is available in its entirety in machine-readable form. The first six entires are shown here for guidance regarding its format and content.}
\end{deluxetable*}

\subsection{Hertzsprung-Russell Diagram}\label{sec:HR}
\begin{figure*}
	\centering
	\includegraphics[scale=0.8]{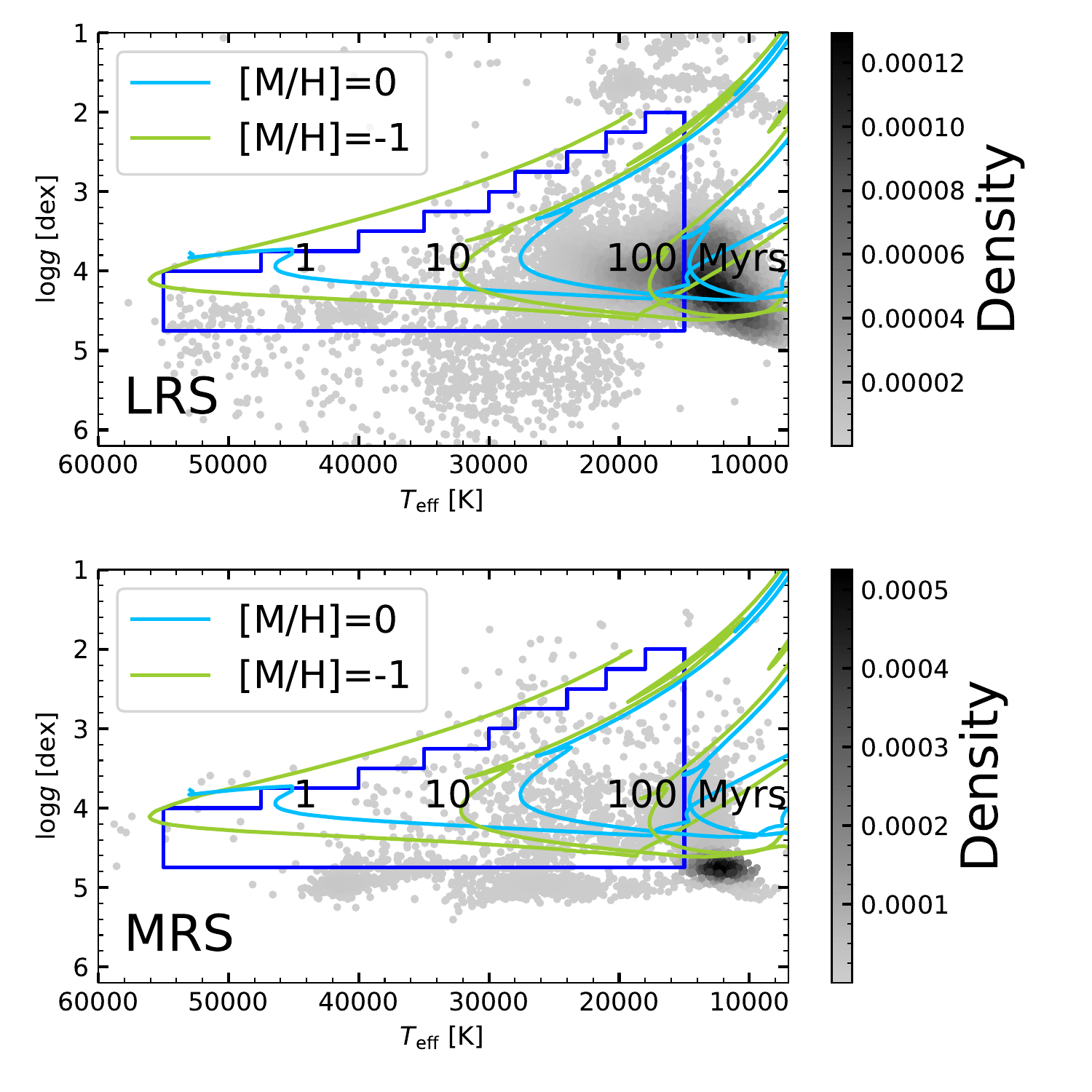}
    \caption{The distribution of predicted stellar labels of early-type stars for both LAMOST-LRS (upper panel) and LAMOST-MRS (bottom panel) spectra is shown in the ($T_\mathrm{eff}$, $\log{g}$) plane. The vertical color bars on both panels represent the number density of the stars. The blue box represents the restricted parameter range of $T_\mathrm{eff}$ and $\log{g}$ values selected from the TLUSTY model grids. The green and blue lines represent PARSEC the theoretical isochrones tracks with metallicity values of $[M/H] = -1.0$ dex and $[M/H] = 0.0$ dex, respectively. A set of tracks with ages of 1, 10, and 100 Myrs are included in the figure.}
\label{fig:LAMOST parameter density}
\end{figure*}

In Fig.~\ref{fig:LAMOST parameter density} we show the density distributions of predicted $T_\mathrm{eff}$ and $\log{g}$ values for both LAMOST-LRS and LAMOST-MRS spectra. The vertical color bar on the right represents the number density of the stars, and the blue box represents the restricted parameter ranges of $T_\mathrm{eff}$ and $\log{g}$ selected from the TLUSTY model grids (see \ref{sec:train set}). We derived the isochrones from the Padova and Trieste Stellar Evolutionary Code (PARSEC) \citep{2000Hurleyisoyear,2012Bressan}, and overplotted them with the ages of 1, 10, and 100 Myrs on the figure. Two sets of isochrones with values of $[M/H] = -1.0$ dex (green lines) and $[M/H] = 0.0$ dex (blue lines) are shown for each of the age track. The theoretical isochrones are well fitted with the predicted stellar labels for the observed spectra as given by the {\tt SLAM}.

We see in Fig.~\ref{fig:LAMOST parameter density} that a large fraction of the stars in the sample with predicted stellar labels are located outside the restricted parameter range of our training set. Based upon their location on the plot, they can be divided into three groups, and we briefly discuss their features below. The first group of stars shows an estimated $T_\mathrm{eff} < 15,000$ K (with spectral-type later than B5). This suggests that late B-type stars dominate the population of our sample of LAMOST early-type stars, and such finding is consistent with the result suggested by the initial mass function, i.e.\ the population of late-type stars shows an increasing trend towards the small stellar mass. In upper panel of Fig.~\ref{fig:LAMOST parameter density}, a second group of stars consisting of 363 LAMOST-LRS spectra showing predicted values of $\log{g} < 2.0$ cm s$^{-2}$. We visually inspected the spectra of these stars and found that they are either featureless or dominated by the presence of the H$\alpha$ emission profile. We caution that such features appearing in the spectra might lead to an underestimation of the true $\log{g}$ values of the observations. 
The last group of stars is found with predicted values of $\log{g} > 4.75$ cm s$^{-2}$, and they are likely subdwarf-type stars, which have typical $\log{g}$ values in the range of 5.1 to 6.4 cm s$^{-2}$ \citep{2009Hebersdb}.

\subsection{Comparing to high-resolution spectra}\label{sec:compare High ST}
\begin{figure*}
\begin{center}
	\centering
	\includegraphics[scale=0.4]{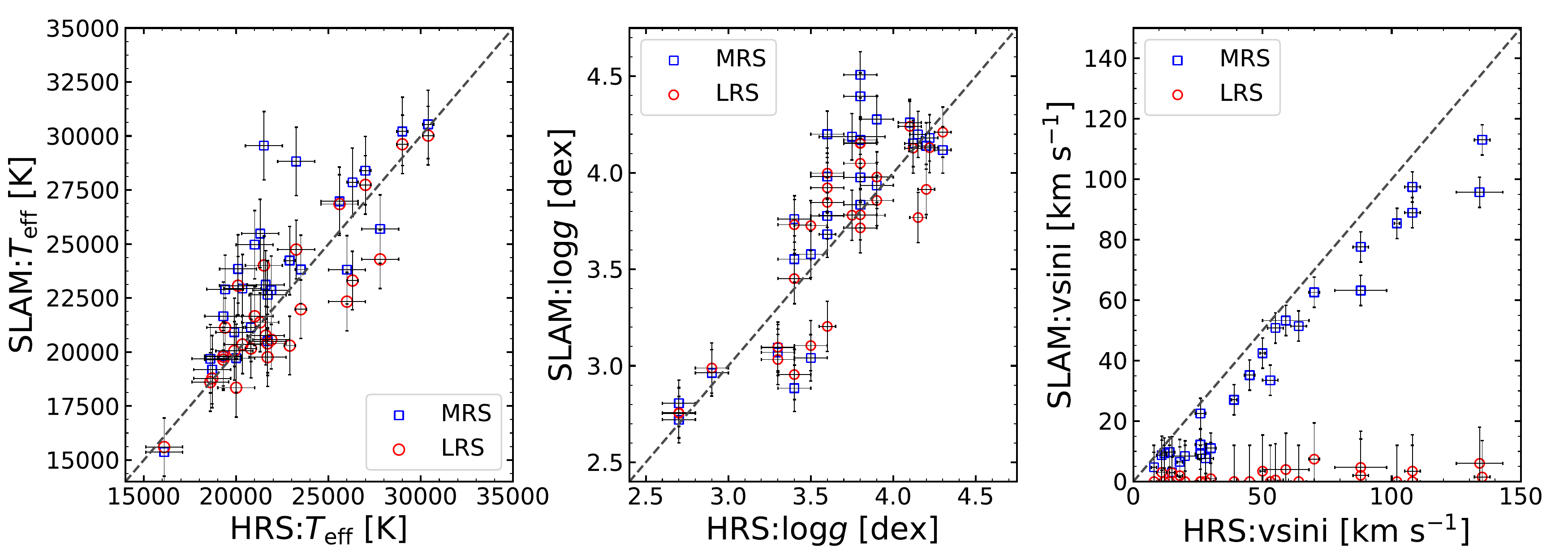}
	\caption{Comparisons of predicted stellar labels of $T_\mathrm{eff}$, $\log{g}$, and $V\sin{i}$ of 28 early-type stars obtained from the {\tt SLAM} to the pre-labeled values as given by \cite{hunter2009}, \cite{nieva2012}, and \cite{mvevoy2017}.}
	\label{fig:compare HST}
\end{center} 
\end{figure*}

.
We also predict the stellar labels for a sample of pre-labeled high-resolution spectra (HRS) from publications to verify the consistency of our training set. We select a sample of 28 early-type stars with pre-estimated stellar labels from works of \cite{trundle2007}, \cite{nieva2012} and \cite{mvevoy2017}. In order to directly compare the results of these archived HRS spectra to those of the LAMOST-LRS and LAMOST-MRS data, we first degraded the resolution of the HRS spectra down to resolutions of $R\sim1,800$ (HRS-LRS) and $R\sim7,500$ (HRS-MRS). We then applied the {\tt SLAM} to both of the degraded HRS-LRS and HRS-MRS 
spectra to predict their stellar labels values, including the $T_\mathrm{eff}$, $\log{g}$, and $v\sin{i}$. A comparison of our predicted stellar labels using the {\tt SLAM} to the pre-labeled values obtained from publications is shown in Fig.~\ref{fig:compare HST}. The left and middle panels indicate that the predicted $T_\mathrm{eff}$ and $\log{g}$ values from the HRS-LRS spectra (red circles) agree with the published estimates of HRS within the uncertainty range. These two parameters are better constrained comparing to those of the HRS-MRS spectra (blue squares). The right panel shows that the predicted $v\sin{i}$ values of the HRS-MRS spectra are sightly smaller than values given by the HRS, while those estimates from HRS-LRS spectra display significant deviation from the HRS results. Detailed line features from the HRS spectra likely were flattened during the spectral degradation process, resulting in a smaller estimation for the $v\sin{i}$.

In Section~\ref{sec:err SNR}, we discuss the errors of the predicted stellar labels given by {\tt SLAM}. Assuming a SNR$ = 100$ is given for input TLUSTY model spectra, based on the distribution of {\tt SLAM} errors of stellar labels as a function of SNR values shown in Fig.~\ref{fig:err SNR}, we would expect to obtain estimated errors of $\sigma(T_\mathrm{eff}) = 327 $K, $\sigma(\log{g}) = 0.03$ dex, and $\sigma(v\sin{i}) = 4\, \rm km\,s^{-1}$ for TLUSTY-LRS spectra. For training set consisting of model spectra of TLUSTY-MRS, we obtained errors of $\sigma(T_\mathrm{eff}) = 562 $K, $\sigma(\log{g}) = 0.04$ dex, and $\sigma(v\sin{i}) = 1\, \rm km\,s^{-1}$. We caution that these estimation from the model spectra may underestimate the true values of observational data. We thus calculated the standard deviation as realistic errors between the predicted stellar label and the pre-labeled published values from the HRS for each stellar parameter as the realistic errors of the labels. We then arrived to realistic error estimates of $\sigma(T_\mathrm{eff}) = 2,185 $K, $\sigma(\log{g}) = 0.29$ dex, and $\sigma(v\sin{i}) = 11\, \rm km\,s^{-1}$ for HRS-MRS spectra, and $\sigma(T_\mathrm{eff}) = 1,642 $K, $\sigma(\log{g}) = 0.25$ dex, and $\sigma(v\sin{i}) = 42\, \rm km\,s^{-1}$ for HRS-LRS spectra. These error estimates likely reflect the true uncertainties of stellar labels derived from the {\tt SLAM}.

\subsection{Comparing the performance of LRS and MRS from SLAM}\label{sec:compare MRS LRS}
\begin{figure*}
\begin{center}
	\centering
	\includegraphics[scale=0.6]{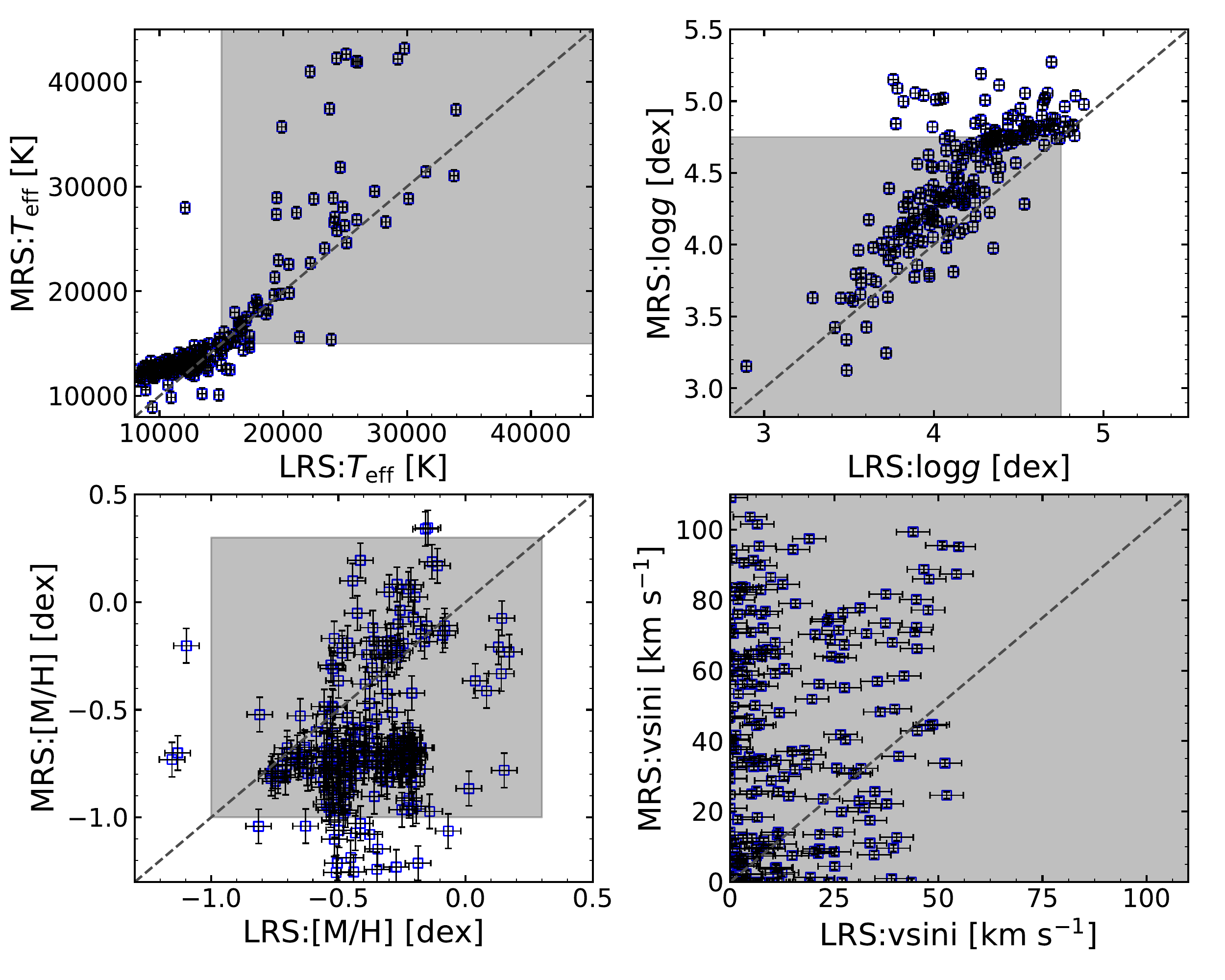}
	\caption{Comparison of predicted stellar labels obtained from the training sets consisting of LAMOST-LRS (x-axis) to LAMOST-MRS (y-axis) spectra. The grey region represents the restricted parameter range selected from the TLUSTY model grids.}
	\label{fig:compare mrslrs}
\end{center} 
\end{figure*}

In order to assess the dependence of predicted stellar labels of the spectral resolution using the {\tt SLAM}, we cross-matched the early-type stars listed in \cite{2019LiuZhicun} (LAMOST-LRS) with those of Paper~I (LAMOST-MRS). A sample of 229 common stars with an SNR value greater than 100 was found. In Fig~\ref{fig:compare mrslrs}, we compare the predicted stellar labels from the LAMOST-MRS spectra (y-axis) to those from the LAMOST-LRS (x-axis). The grey region on each panel of Fig~\ref{fig:compare mrslrs} represents the restricted parameter range of the stellar parameters selected from the TLUSTY model grids (See Sec.~\ref{sec:train set}). A small sample of stars that are located outside the regions (See Sec.~\ref{sec:HR}). We find that most of the predicted stellar labels of $T_\mathrm{eff}$, $\log{g}$ and $v\sin{i}$ values derived from the LAMOST-MRS spectra show higher values than those from the LAMOST-LRS.

Based upon the results shown from the validation tests discussion in Sec.~\ref{sec:consistency check}, \ref{sec:err SNR}, and \ref{sec:compare High ST}, we conclude that the stellar labels of $T_\mathrm{eff}$, $\log{g}$, and $[M/H]$ obtained from LAMOST-LRS spectra are better constrained comparing to those of LAMOST-MRS spectra for early-type stars. Similar results are concluded from the work of \cite{2020zhangboMRS}, who
performed the {\tt SLAM} model to predict the stellar labels for F-, G-, and K-type stars using spectra from LAMOST. It is likely that the LAMOST-LRS spectra have a broad wavelength coverage of 3,900\,\AA\ to 7,000\,\AA\ comparing to that of LAMOST-MRS spectra, more Balmer and He I line series are included in the spectra resulting in a better constraint on $T_\mathrm{eff}$ and $\log{g}$. Due to the higher resolution of LAMOST-MRS spectra with an R$\sim$7,500 comparing to an R$\sim$1,800 for LAMOST-LRS spectra, more detailed spectral information, such as the outline and shape of broad absorption line profiles are encoded in the MRS spectra. This results in a better estimation of the projected rotational velocity. We suspect that the $v\sin{i}$ is sensitive to the spectral resolution.

\section{Discussion}\label{sec:Discussion}
\begin{figure*}
\begin{center}
	\centering
	\includegraphics[scale=0.6]{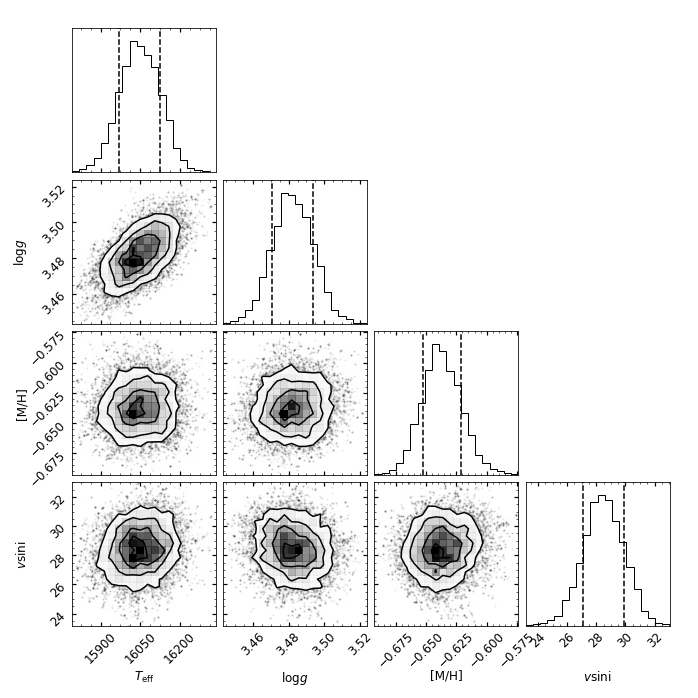}
	\caption{Posterior distributions of predicted $T_\mathrm{eff}$, $\log{g}$, [M/H] and $v\sin{i}$. The contours from the inside out enclose the 68 and 95 and 99 percentiles of the total probability in the six off-diagonal panels. The two vertical dashed lines in the four diagonal panels represent the lower and upper errors of the distribution at the 16 and 84 percentile, respectively.}\label{fig:MCMC}
\end{center} 
\end{figure*}

The stellar labels of massive stars are enclosed in their spectra, and the stellar parameters are often interrelated. For example, for B-type stars, the neutral helium and hydrogen lines are popular lines to estimate effective temperature, while the hydrogen lines are sensitive to surface gravity \cite{1999Dufton}. Traditionally, the stellar labels are determined individually through the minimization technique by comparing observed spectra to model spectra, while the {\tt SLAM} predicts the stellar labels of input spectra simultaneously.

In order to verify the potential degeneracy existing among the predicted stellar labels obtained from the {\tt SLAM}, we applied an MCMC simulation to inspect the posterior distribution of the stellar labels. We show the results obtained from a LAMOST-MRS spectrum with an observational ID of OBSID 682701103 observed on the night of MJD 58470.9540 in Fig.~\ref{fig:MCMC}. In the six off-diagonal panels, the contours with 68, 95, and 99 percent confidence regions are plotted in solid black lines from the inside out. In the four diagonal panels, the two vertical dashed lines represent the lower and upper errors of the distribution for the individual stellar label at the 16 and 84 percentile, respectively. The distribution indicates that a weak correlation is found between the $T_\mathrm{eff}$ and $\log{g}$. However, the variation of the determined stellar labels is within the uncertainty range estimated from the {\tt SLAM}, and we thus consider the predicted results are still acceptable.

\section{Conclusion}\label{sec:Conclusion}
Massive stars are important contributors to many astronomical mechanisms. Determining their fundamental physical parameters is vital to understand the evolutionary scenario of the massive stars. Hitherto, the estimate of stellar parameters for such stars is restricted to small collections of spectral observations. Estimating the stellar labels using a comprehensive and consistent approach for a large sample of early-type stars is missing. Motivated by the recent lease of a large number of spectra for early-type stars from the LAMOST database, we thus conduct this study to predict stellar labels for more than 20,000 early-type stars identified from the database.

We adopted a synthetic non-LTE atmospheric spectral library, TLUSTY, to generate training sets of TLUSTY-LRS and TLUSTY-MRS model spectra through linear interpolation. The model spectra encompass a wide parameter range to include spectral features of early-type stars over the wavelength regime covered by the LAMOST observations. We then applied a 5-fold Cross-Validation technique to validate the performance of the training sets. By investigating the distribution of {\tt SLAM} errors of predicted stellar labels as a function of SNR for input training set with different sample sizes, we determined a sample size of 1,000 spectra for the TLUSTY-LRS train set and 5,000 spectra for the TLUSTY-MRS train set. We predicted the stellar labels of identified early-type stars from the LAMOST database using a sample of 3,931 stars from the LAMOST-LRS and 578 stars from the LAMOST-MRS.

In order to estimate the realistic errors of predicted stellar labels obtained from the {\tt SLAM}, we collected the high-resolution spectra of 28 stars from publications. By downgrading the resolution to match the ones of the observed LAMOST-LRS and LAMOST-MRS spectra, we then predicted the stellar labels of these spectra and determined the uncertainties of the associated label by computing the standard deviation between the predicted values from the {\tt SLAM} and literature values. We then arrived to realistic error estimates of $\sigma(T_\mathrm{eff}) = 2,185 $K, $\sigma(\log{g}) = 0.29$ dex, and $\sigma(v\sin{i}) = 11\, \rm km\,s^{-1}$ for HRS-MRS spectra, and $\sigma(T_\mathrm{eff}) = 1,642 $K, $\sigma(\log{g}) = 0.25$ dex, and $\sigma(v\sin{i}) = 42\, \rm km\,s^{-1}$ for HRS-LRS spectra. Predicted stellar labels of $T_\mathrm{eff}$ and $\log{g}$ obtained from the MRS spectra are better constrained comparing to those from the LRS spectra, and this is likely due to the broader wavelength coverage of the LRS spectra. However, MRS spectra are more sensitive to constrain the $v\sin{i}$ values due to the inclusion of detailed line profile features as a result of higher resolutions. 

We have demonstrated the applicability of applying the data-driven technique, {\tt SLAM}, to predict stellar labels for a large set of observations for early-type stars. This technique is promising to predict stellar labels for other types of stars.             

\acknowledgments
This work is supported by the Natural Science Foundation of China (Nos.\ 11733008, 12090040, 12090043, 11521303), by Yunnan province, by the National Ten-thousand talents program. C.L.\ acknowledges National Key R$\&$D Program of China No.\ 2019YFA0405500 and the NSFC with grant No.\ 11835057.

Guoshoujing Telescope (the Large Sky Area Multi-Object Fiber Spectroscopic Telescope LAMOST) is a National Major Scientific Project built by the Chinese Academy of Sciences. Funding for the project has been provided by the National Development and Reform Commission. LAMOST is operated and managed by the National Astronomical Observatories, Chinese Academy of Sciences.

This work is also supported by the Key Research Program of Frontier Sciences, CAS, Grant No.\ QYZDY-SSW-SLH007.

\bibliography{sample63}{}

\bibliographystyle{aasjournal}
\end{document}